\documentclass[11pt]{article}
\usepackage[left=2.54cm,top=2.54cm,right=2.54cm,bottom=2.54cm]{geometry}

\input{macros/preamble.tex}

\usepackage{rotating}
\usepackage{xcolor}

\begin{document}
\title{\huge Latent Factor Point Processes for Patient Representation in Electronic Health Records}
\author{Parker Knight\thanks{Department of Biostatistics, Harvard T.H. Chan School of Public Health. Contact: \texttt{pknight@g.harvard.edu}.}, Doudou Zhou\thanks{Department of Statistics and Data Science, National University of Singapore.}, Zongqi Xia\thanks{Department of Neurology, University of Pittsburgh.}, Tianxi Cai\thanks{Department of Biostatistics, Harvard T.H. Chan School of Public Health.}, Junwei Lu\thanks{Department of Biostatistics, Harvard T.H. Chan School of Public Health.}}
\date{\today}
\maketitle



\begin{abstract}
Electronic health records (EHR) contain valuable longitudinal patient-level information, yet most statistical methods reduce the irregular timing of EHR codes into simple counts, thereby discarding rich temporal structure. Existing temporal models often impose restrictive parametric assumptions or are tailored to code-level rather than patient-level tasks. We propose the latent factor point process model, which represents code occurrences as a high-dimensional point process whose conditional intensity is driven by a low-dimensional latent Poisson process. This low-rank structure reflects the clinical reality that thousands of codes are governed by a small number of underlying disease processes, while enabling statistically efficient estimation in high dimensions. Building on this model, we introduce the Fourier-Eigen embedding, a patient representation constructed from the spectral density matrix of the observed process. We establish theoretical guarantees showing that these embeddings efficiently capture subgroup-specific temporal patterns for downstream classification and clustering. Simulations and an application to an Alzheimer’s disease EHR cohort demonstrate the practical advantages of our approach in uncovering clinically meaningful heterogeneity.

\end{abstract}

{\textbf{Keywords:} High-dimensional point process, EHR, Spectral methods, Embedding.}

\section{Introduction}

The growing accessibility of electronic health record (EHR) datasets has beget new paradigms of public health research and precision medicine. Recent studies have demonstrated the utility of EHRs for building risk prediction models \citep{goldstein2016opportunities, rasmy2021med, zhang2019metapred, corey2018development}, discovering clinically meaningful patient subgroups \citep{guo2023assessing, jiang2024soft, xu2020identifying, mullin2021longitudinal}, and designing optimal individual treatment plans \citep{wang2016learning,reddy2019predicting,zhou2024federated,chen2025federated}, all of which carry substantial potential to advance clinical decision making \citep{abul2019personalized, giordano2021accessing}. These successes have spurred the development of new statistical methods designed to address challenges unique to EHR analysis, such as multi-institutional integration \citep{duan2020learning, zhou2024federated, xiong2023federated, tong2022distributed, li2024efficient, hu2024collaborative}, semi-supervised learning due to limited gold-standard labels \citep{gronsbell2018semi, cai2025semi, zhang2023double, wang2025semi, nogues2022weakly, zhang2022prior, hou2023surrogate}, and robustness under patient distribution shifts \citep{wang2022survmaximin, zhan2024transfer, knight2025fast, lee2023stable, wang2023distributionally, xiong2023distributionally, luo2022deep, li2024transport}. 

However, a fundamental aspect of structured EHR data remains underutilized: its \textit{high-dimensional temporal structure}. Structured EHRs consist of event times of thousands of billing codes, representing diagnoses, prescriptions, laboratory results, and procedures. Most existing methods collapse event times into vectors of counts over fixed observation windows, and even those that account for time typically do so using only crude temporal summaries. Such strategies do not effectively capture temporal information that may be strongly associated with clinical outcomes \citep{yuan2021temporal}. This has inspired a growing field of interdisciplinary research aiming to design models and methods for integrating temporal information into the analysis of EHRs. One recent line of work endeavors to model EHRs as a high-dimensional Hawkes process \citep{hawkes1971spectra, choi2015constructing, bao2017hawkes, alaa2017learning, sun2024learning, zhao2025balancing}, which enables researchers to learn time-informed interaction networks between large numbers of codes by estimating a high-dimensional conditional intensity function. A parallel literature leverages the temporal information in EHRs to learn medical concept embeddings using contemporary tools from machine learning (ML) \citep{si2021deep}, including recurrent neural networks \citep{ruan2019representation, rasmy2022recurrent, zhu2023prediction, al2024ta}, transformer architectures \citep{li2020behrt, choi2015constructing, yang2023transformehr, yu2025time}, and foundation-scale large language models \citep{yang2022large, steinberg2021language, agrawal2022large, li2024scoping}. These Hawkes process and ML based methods are able to learn complex relationships between medical concepts over time from the health record, but do so without providing theoretical guarantees, leaving the landscape of statistical theory for time-informed EHR methods underexplored. Recent work has begun to address this by adapting statistical models from text and sequence analysis \citep{arora2016latent}, in particular log-linear hidden Markov models \citep{lu2023knowledge, xu2023inference, zhou2023multi}, which capture temporal dependence between codes with simple co-occurrence statistics for building medical concept embeddings and knowledge graphs \citep{beam2020clinical, hong2021clinical, gan2025arch}. 
The authors of these works are able to prove precise and informative theorems demonstrating the statistical properties of their methods. However, they do so by imposing restrictive parametric assumptions on the code occurrence probabilities and assuming discrete and evenly spaced event times, at odds with the irregularity of real EHR data. Most importantly, all of these methods are designed to capture concept-level relationships, not patient-level representations, which severely detracts from their utility in downstream tasks of high clinical importance, such as risk prediction and unsupervised patient stratification.

To bridge this gap, we propose a new model for temporal EHR data: the \textit{latent factor point process model}. In this model, the code event times are represented as a high-dimensional point process whose conditional intensity is driven by a low-dimensional, unobserved latent factor Poisson process. This structure is inspired by clinical reality: although thousands of heterogeneous codes are recorded in EHRs, their co-occurrence is often driven by a small number of underlying medical processes (e.g., disease progression, comorbidity conditions, treatment response). In addition to clinical interpretability, the low-rank structure of our proposed model is statistically essential for estimating parameters of interest in high dimensions.

Motivated by patient-level tasks such as classification and clustering, we assume that patients belong to distinct subgroups and that the latent intensity vector varies systematically across these groups. In contrast to the linear Hawkes process and other parametric point process models \citep{daley2003introduction, bacry2020sparse}, we do not impose specific functional forms on transfer functions, allowing the event probabilities to flexibly evolve with the latent history. We show that the cross-covariance matrix of the observed codes inherits a low-rank structure, and that the spectral properties of this matrix differ systematically across patient subgroups. This motivates our novel Fourier-Eigen embedding, a time-informed patient representation constructed from the spectral density matrix of the observed process.

We provide a rigorous theoretical analysis showing that Fourier-Eigen embeddings serve as efficient patient representations for downstream tasks, including empirical risk minimization in classification and subgroup discovery in clustering. We supplement our analysis with simulation studies and an application to an Alzheimer’s disease cohort, a clinically important setting where capturing temporal progression is critical for understanding heterogeneous disease trajectories.

\subsection{Related literature}

Our work builds on recent developments in the high-dimensional point process literature \citep{hansen2015lasso, bacry2016first, hawkes2018hawkes}. The latent factor point process model that we propose is inspired by the multivariate temporal point process regression model of \citet{tang2023multivariate}, where the conditional intensity of an outcome process is driven entirely by a high-dimensional predictor process. In our setting, the predictor process corresponds to the latent factor process and the outcome process to the observed code occurrence process. A key difference is that our latent factor process is unobserved, which makes the estimation strategy of \citet{tang2023multivariate} inapplicable. \cite{wang2021causal} also considers a high-dimensional point process model in which a subset of the component processes are unobserved yet still affect the conditional intensity of the observed processes. 
Under a linear assumption, they provide a deconfounded estimator of the intensity coefficients, which is consistent when the bias from unmeasured components is negligible. In contrast, our model assumes that the conditional intensity of the observed process is entirely driven by an unobserved process, and we do not require the intensity to follow a particular functional form.

The construction of our Fourier-Eigen embeddings builds on spectral methods for point processes. The spectral density matrix, defined as the Fourier transform of the cross-covariance matrix, has been widely used in classical work on point processes \citep{bartlett1963spectral, hawkes1971spectra, bacry2016first} and time series in general \citep{koopmans1995spectral}. For example, \citet{bacry2016first} derive Wiener–Hopf equations for a multivariate Hawkes process, which they solve in the Laplace domain to obtain a nonparametric estimator of the conditional intensity. More recently, \citet{pinkney2024regularised} study the multi-taper periodogram for estimating high-dimensional spectral density matrices and propose a regularized inverse spectral estimator, while \citet{hellstern25spectral} develop an estimator for the difference between two spectral density matrices in differential network analysis. These advances highlight the utility of spectral approaches for extracting structure from complex stochastic processes, and we adapt this perspective to define low-dimensional patient embeddings.

Finally, our work contributes to the growing literature on statistical models for temporal EHR data. The seminal work of \citet{arora2016latent} analyzed a log-linear hidden Markov model for text generation, showing that the pointwise mutual information matrix captures latent embeddings of words up to a tolerable error. This idea has since been adapted to EHR analysis: for instance, \citet{lu2023knowledge} allow the code embeddings to admit a block-wise covariance structure for estimating structured knowledge graphs of medical concepts, while \citet{xu2023inference} provide edge-wise inference on such graphs. Related extensions include multi-institutional adaptations \citep{zhou2023multi} and code-level clinical knowledge extraction \citep{beam2020clinical, hong2021clinical, gan2025arch}. More recently, \citet{yu2025time} propose a marked point process model with time-aware attention to learn code embeddings. These works are primarily designed for capturing relationships between codes. By contrast, our goal is to develop representations tailored to patient-level tasks such as classification and clustering, while flexibly incorporating the temporal structure of EHR data.

\subsection{Notation}

For two real numbers $a, b \in \RR$, we let $a \land b = \min(a, b)$ and $a \lor b = \max(a, b)$. For a matrix $\mathbf{A} \in \CC^{n \times m}$, we let $\{\sigma_i(\mathbf{A})\}_{i = 1}^{n \land m}$ denote its singular values in decreasing order. We use $\|\mathbf{A}\|_F = \sqrt{\sum_{i,j}|A_{ij}|^2}$ and $\| \mathbf{A}\|_{\text{op}} = \sigma_1(\mathbf{A})$ to denote the Frobenius and operator norms of $\mathbf{A}$ respectively. We will use $\mathbf{A}^{\dagger}$ to denote the conjugate transpose of $\mathbf{A}$, and we say that the matrix $\mathbf{A}$ is Hermitian if $\mathbf{A} = \mathbf{A}^{\dagger}$. For two functions $f, g: \RR \to \RR$, we write their  convolution as $(f*g)(t)=\int f(\tau)g(t - \tau)\D \tau$. If $\mathbf{A}(t) \in \RR^{n\times m}$ and $\mathbf{B}(t) \in \RR^{m \times q}$ are matrix-valued functions with entries $A_{ij}(t)$ and $B_{j\ell}(t)$, we define $\big(\mathbf{A} \star \mathbf{B}\big)(t)$ as the matrix with entries $(A \star B)_{i\ell}(t) = \sum_{j = 1}^{m}(A_{ij} * B_{j\ell})(t)$. For an absolutely integrable function $f: \RR \to \RR$, we let its Fourier transform evaluated at frequency $\xi$ be $\sF \{f\}(\xi) = \int_{\RR} f(x)\exp(-i2\pi\xi x)dx$. If $\mathbf{A}(t) \in \RR^{n\times m}$ has integrable elements $A_{ij}(t)$, then $\sF\{\mathbf{A}(t)\}(\xi)$ denote the matrix with entries $\sF\{A_{ij}\}(\xi)$.

A univariate temporal point process $N$ is a stochastic process on $\RR^+$ with event times $\{t_\ell : \ell=1,2,\dots\}$. We write $N(A) = \sum_{\ell} \one(t_\ell \in A)$ for $A \subseteq \RR^+$. and $dN(t) = N([t, t+dt])$ for a small increment $dt$. The mean intensity of $N$ is
$\Lambda(t) = \lim_{dt \rightarrow 0} \EE[dN(t)]/dt$. We say $N$ is stationary if the distribution of $dN(t)$ does not depend on $t$. The conditional intensity function of $N$ is defined by 
\[\lambda^N(t)dt = \PP(dN(t) = 1 | \cH_t),
\]
where $\cH_t$ is the history of $N$ up to time $t$ (i.e. the $\sigma$-algebra generated by ${N(u):0<u<t}$).

We define a multivariate point process as $\bN = (N_1, ..., N_d)^{\top} \in \RR^d$ with each element being an univariate point processes. If $\bN$ is stationary, we define its cross-covariance matrix as 
\[V_{ij}(\tau) = \frac{\EE[dN_i(t)dN_j(t - \tau)]}{dtd(t-\tau)} - \Lambda_i\Lambda_j - \delta_{ij}(\tau)\Lambda_i, i,j = 1,\ldots,d,\]
where $\tau \in \RR$, $\Lambda_i$ is the mean intensity of $N_i$, and $\delta_{ij}(\tau) = \one(i = j)\delta(\tau)$ with $\delta(\cdot)$ the Dirac delta function. 

Finally, for $\mu \in \RR^k$, we write $\bM \sim \text{PoisProc}(\mu, T)$ if $\bM$ is a homogeneous Poisson process on $\RR^k$ with intensity vector $\mu$ observed over $[0,T]$. Throughout, $c_1,c_2,c_3,c_4,C,C'$ denote positive constants whose values may vary across uses.


\section{Model and methods}

\subsection{Latent factor point process model}


We let $\bN$ denote a $d$-dimensional point process, $\bX \in \RR^p$ a vector of baseline covariates, $\bM$ an unobserved $k$-dimensional point process whose meaning will be made clear in the sequel, and $G$ is a class indicator that takes values in a finite set $\cG$. The full dataset consists of $n$ i.i.d. quartets $\cD = \{(\bN_i, \bX_i, \bM_i, G_i)\}^n_{i = 1}$, but we only assume that we will have access to either $\cO_{\text{sup}} = \{(\bN_i, \bX_i, G_i)\}^n_{i = 1}$ or $\cO_{\text{unsup}} = \{(\bN_i, \bX_i)\}^n_{i = 1}$; in either case, we observe the process $\bN_i$ from time $0$ to $T_i > 0$ for each $i \in \{1, \ldots, n\}$. The subscripts ``sup'' and ``unsup'' refer to the supervised and unsupervised settings, respectively, and indicate whether or not the labels $\{G_i\}_{i=1}^n$ are observed. In the context of EHR data, $\bN_i$ denotes the occurrence times for $d$ EHR codes observed for patient $i$, and $G_i$ may denote a disease status indicator. Our overarching goal is to accurately distinguish the patients belonging to the classes in $\cG$ using the observed data, which may or may not contain labels $\{ G_i\}_{i = 1}^n$. Our first step in this direction is to articulate a new generative model for high-dimensional point processes that simultaneously describes the correlation structure between codes that we observe in real-world EHR data and captures differences between the subgroups of patients. We will impose the following model on the process $\bN_i$:
\begin{align}
    &G_i \sim \pi(\cG), \label{eq:g-pi}\\
    &\bM_i \, | \, G_i = g \sim \text{PoisProc}(\mu^{(g)}, T_i),\label{eq:mean-diff} \\
    &\lambda^{N_{ij}}(t) = \nu_j(\bX_i) + \sum_{\ell = 1}^k\big(\omega_{j\ell} * dM_{i\ell}\big)(t).\label{eq:lfpp}
\end{align}
Here, $\pi$ is an arbitrary distribution on $\cG$ that assigns nonzero mass to each $g \in \cG$. As is standard in the point process literature \citep{hansen2015lasso}, we state our model for $\bN_i$ in (\ref{eq:lfpp}) in terms of the conditional intensity function $\lambda^{N_{ij}}$. We let $\bM_i$ denote an $k$-dimensional unobserved, latent homogeneous Poisson process whose intensity vector $\mu^{(g)}$ is determined by the latent class $G_i$.  
The choice of a homogeneous Poisson latent driver is a parsimonious assumption that induces a low-rank second-order structure while keeping estimation feasible in high dimensions; extensions to inhomogeneous or self-exciting latent processes can introduce additional analytical challenges, and we leave their exploration to future work. 

In (\ref{eq:lfpp}), $\nu_j : \RR^p \rightarrow \RR$ is the background intensity of feature $j$ which is a function of baseline covariates, and $\omega_{j\ell} : \RR^{+} \mapsto \RR^{+}$ is a transfer function that controls the extent to which the history of the unobserved process $M_{i\ell}$ affects the intensity of feature $j$. We can gain intuition for the role that $\bM_i$ plays in the conditional intensity of $\bN_i$ by unwinding the convolution and the Riemann-Stieltjes integral in (\ref{eq:lfpp}) as follows:
\begin{align}
    \lambda^{N_{ij}}(t) &= \nu_j(\bX_i) + \sum_{\ell = 1}^k\big(\omega_{j\ell} * dM_{i\ell}\big)(t) \\
    &= \nu_j(\bX_i) + \sum_{\ell = 1}^k\int_0^t\omega_{j\ell}(t - u)d{M_{i\ell}}(u) \\
    &= \nu_j(\bX_i) + \sum_{\ell = 1}^k \sum_{ \text{$M_{i\ell}$ occurs at time $u < t$}}\omega_{j\ell}(t - u).\label{eq:lfpp-intuition}
\end{align}
In (\ref{eq:lfpp-intuition}), we can see that each occurrence of $M_{i\ell}$ triggers an increase in the probability of $N_{ij}$ occurring through the transfer function $\omega_{j\ell}$. Analogous to the structure of latent factor models used in the time series analysis literature \citep{lam2011estimation, molenaar1985dynamic, bartholomew2011latent}, the process $\bM_i$ acts as the latent factor process that drives the occurrences observed in $\bN_i$. For this reason, we say that $\bN_i$ follows the \textit{latent factor point process model} with latent factors $\bM_i$. We emphasize that the transfer functions $\{\omega_{j\ell}\}$ are not dependent on the index $i$, and are shared in common across all patients. The history of patient $i$ affects $\lambda^{N_{ij}}$ only through $\bM_i$, whose intensity is determined by $G_i$.

Model (\ref{eq:lfpp}) is inspired by the structure of electronic health record data. The occurrence of EHR codes is caused not by the prior occurrence of other codes, but by real-world medical events that are represented in the health record only by proxy. For instance, a myocardial infraction (a heart attack) is denoted in the health record by several different ICD-10 codes \citep{rockenschaub2020data}, which will in turn co-occur with codes denoting medical procedures and treatments for patients who have experienced a heart attack. The latent process $\bM_i$ can be interpreted as these real-world events whose occurrence is observed only through the ambient process $\bN_i$. In light of this interpretation, equation (\ref{eq:mean-diff}) states that the rates of the real-world latent events differ systematically between the classes of patients, while the transferring functions that map these events to the health record are shared across all patients. The intuition for the role of $\bM_i$ as representing real-world events also motivates our specification of the latent process as a Poisson process rather than, say, a Gaussian process as used in \citet{lu2023knowledge} and  \citet{xu2023inference}.

Furthermore, Model (\ref{eq:lfpp}) grants a low rank structure in the cross-covariance matrix of the process $\bN_i$ that interpretably encodes heterogeneity between the groups of patients. The following proposition, which is a special case of Proposition 1 in \citep{tang2023multivariate}, describes this formally. We provide a proof in the supplement for completeness.
\begin{proposition}\label{prop:lowrank}
    Suppose that $\bN_i \in \RR^d$ and $\bM_i \in \RR^k$ are stationary point processes that satisfy (\ref{eq:lfpp}). Let $\mu_i \in \RR^k$ denote the mean intensity of the process $\bM_i$ with entries  $\mu_{i\ell} = \lim_{dt \rightarrow 0}\EE[dM_{i\ell}(t)]/dt$, and assume that $M_{i\ell}$ and $M_{i\ell'}$ are independent for $\ell \neq \ell'$. Then the elements of the instantaneous cross-covariance matrix of $\bN_i(t)$ at lag $\tau \neq 0$ are given by 
    \begin{equation}\label{eq:v-elem}
        V^{\bN_i}_{jj'}(\tau) = \sum_{\ell = 1}^k\mu_{i\ell}\cdot (\omega_{j\ell} * \omega_{j'\ell})(\tau).
    \end{equation}
    Defining $t \mapsto \mathbf{D}_i(t)$ as the constant function that returns $\text{diag}(\mu_i)$, we can thus express $V^{\bN_i}$ as 
    \begin{equation}\label{eq:v-eigen}
    V^{\bN_i}(\tau) = \big(\mathbf{\omega} \star \mathbf{D}_i \star \mathbf{\omega}^{\top}\big)(\tau).
    \end{equation}
\end{proposition}

This low-rank form is the key property that enables us to compress patient trajectories into a small number of spectral features, which underpins the Fourier-Eigen embedding developed below. The form of $V^{\bN_i}$ given in Equation (\ref{eq:v-eigen}) is analogous to its spectral decomposition at $\tau$ with `spectrum' $\mu_i$. We can make this analogy explicit with elementary tools from harmonic analysis. With suitable regularity conditions on the transfer functions $\{\omega_{j\ell}\}$, we invoke the Convolution Theorem \citep{dym1972fourier} to convert the structure given in (\ref{eq:v-eigen}) to that of a standard matrix product using a Fourier transform. This gives us 
\begin{equation}
    \sF\big\{V^{\bN_i}_{jj'}\big\}(\xi) = \sum_{\ell = 1}^k\mu_{i\ell}\cdot \bar{\omega}_{j\ell}(\xi)\cdot \bar{\omega}_{j'\ell}(\xi)
\end{equation}
where we define $\bar{\omega}_{j\ell}(\xi) = \sF\{\omega_{j\ell}\}(\xi)$ as the Fourier transform of $\omega_{j\ell}$ at evaluated at $\xi$. Letting $\mathbf{W}{(\xi)} \in \CC^{d \times k}$ denote the matrix with entries $W{(\xi)}_{j\ell} = \bar{\omega}_{j\ell}(\xi)$, we can thus write 
\begin{equation}
    \sF\big\{V^{\bN_i}\big\}(\xi) = \mathbf{W}{(\xi)}\mathbf{D}_i(\mathbf{W}{(\xi)})^{\top}.
\end{equation}
Although the matrix $\mathbf{D}_i$ is diagonal, this factorization of $\sF\big\{V^{\bN_i}\big\}(\xi)$ does not constitute an eigendecomposition, as we have no guarantee that $\mathbf{W}{(\xi)}(\mathbf{W}{(\xi)})^{\top} = \bI_d$ and $(\mathbf{W}{(\xi)})^{\top}\mathbf{W}{(\xi)} = \bI_k$. However, since $\mathbf{W}(\xi)$ spans the column space of $\sF\big\{V^{\bN_i}\big\}(\xi)$, the leading eigenvalues retain the subgroup-specific information carried by $\mathbf{D}_i$. In particular, the top $k$ eigenvalues of $\sF\big\{V^{\bN_i}\big\}(\xi)$ can be expressed as $\mathbf{H}\mathbf{D}_i$ for a suitable invertible matrix $\mathbf{H}$ as long as $\mathbf{W}{(\xi)}$ has full column rank. Recall that $\mathbf{D}_i = \text{diag}(\mu_i)$. Under (\ref{eq:mean-diff}), this indicates that the eigenvalues of $\sF\big\{V^{\bN_i}\big\}(\xi)$ and $\sF\big\{V^{\bN_{i'}}\big\}(\xi)$ will differ systematically when $G_i \neq G_{i'}$. This explains our use of the term \textit{Fourier-Eigen} in the sequel.

\subsection{Classification and clustering with Fourier-Eigen embeddings}


This exploration of the structure of $V^{\bN_i}$ motivates our proposed procedure for building a classifier or clustering algorithm. We propose to estimate a low-dimensional patient embedding that captures the signal needed to distinguish the classes of patients, and use these embeddings as features in the downstream tasks of classification and clustering. The embeddings are defined as follows. Given a fixed frequency value $\xi_0 \in \RR$, for each $i \in [n]$ we perform the eigendecomposition
\begin{equation}
    \sF\big\{V^{\bN_i}\big\}(\xi_0) = \mathbf{U}_i\Lambda_i\mathbf{U}_i^{-1},
\end{equation}
where $\Lambda_i$ is the diagonal matrix of eigenvalues $\lambda_{i1}, \ldots, \lambda_{id}$. We then define 
\begin{equation}
    f_i = (\lambda_{i1}, \ldots, \lambda_{ik})^{\top}
\end{equation}
as the $k$-dimensional \textit{Fourier-Eigen} embedding of patient $i$. In Section \ref{sec:theory}, we will show formally that if patients $i$ and $i'$ belong to different classes then their embeddings $f_i$ and $f_{i'}$ are well-separated, and therefore a standard classifier trained on the pairs $\{(f_i, G_i)\}^n_{i = 1}$ or a suitable clustering algorithm will be able to perfectly distinguish the classes of patients. Of course, $f_i$ is a population-level quantity, as its construction depends on knowledge of the true cross-covariance matrix $V^{\bN_i}$. We estimate $f_i$ by repeating this Fourier-eigendecomposition procedure with an estimator of $V^{\bN_i}$. Letting $\hat{V}^{\bN_i}$ denote such an estimator, we compute the eigendecomposition
\begin{equation}
    \sF\big\{\hat{V}^{\bN_i}\big\}(\xi_0) = \hat{\mathbf{U}}_i\hat{\Lambda}_i\hat{\mathbf{U}}_i^{-1},
\end{equation}
and define 
\begin{equation}
    \hat{f}_i = (\hat{\lambda}_{i1}, \ldots, \hat{\lambda}_{ik})^{\top},
\end{equation}
where $\hat{\lambda}_{i1}, \ldots, \hat{\lambda}_{id}$ denote the eigenvalues of $\sF\big\{\hat{V}^{\bN_i}\big\}(\xi_0)$. The vector $\hat{f}_i$ will serve as our estimate of the Fourier-Eigen embedding of patient $i$. We note that the vectors $\{\hat{f}_i\}_{i = 1}^n$ obtained from Algorithm \ref{alg:FE} are guaranteed to be in $\RR^k$ as long as $\tau \mapsto \hat{V}^{\bN}(\tau)$ is symmetric in $\tau$, as this ensures that the matrix $\sF\big\{\hat{V}^{\bN_i}\big\}(\xi_0)$ is Hermitian.  We summarize this procedure in Algorithm \ref{alg:FE}. From here, we proceed according to whether or not we observe class labels for each of the $n$ patients. If we observe the labeled dataset $\cO_{\text{sup}}$ we train a classifier on the pairs $\{(\hat{f}_i, G_i)\}^n_{i = 1}$ with Empirical Risk Minimization, computing
\begin{equation}\label{eq:erm}
    \hat{h} = \argmin_{h \in \cH}\frac1n\sum_{i = 1}^n L\big(G_i, h(\hat{f}_i)\big),
\end{equation}
where $\cH$ is a function class on $\RR^k$ with outputs in $\cG$. The loss function $L: \cG^{\otimes 2}\rightarrow \RR$ is chosen by the statistician. Otherwise, if we only have access to $\cO_{\text{unsup}}$, we run K-means clustering with a spectral initialization \citep{kumar2010clustering} on the estimated embeddings $\{\hat{f}_i\}_{i = 1}^n$. We state the classification procedure in Algorithm \ref{alg:classifier} and the clustering procedure in Algorithm \ref{alg:clustering}. To estimate the patient-level cross covariance matrix $V^{\bN_i}$, we use the kernel smoothing estimator proposed by \citep{chen2017multivariate} as described in Algorithm \ref{alg:cov-est}.

\begin{algorithm}
\caption{Estimation of Fourier-Eigen embeddings.}
\label{alg:FE}
\begin{algorithmic}
\State \textbf{input}: Data $\{\bN_i\}_{i = 1}^{n}$, frequency $\xi_0 \in \RR$.
\For{$i$ $\gets$ $1$ to $n$}
\begin{itemize}
    \item Compute $\tau \mapsto \hat{V}^{\bN_i}(\tau)$ with Algorithm \ref{alg:cov-est}.
    \item Factorize $\sF \{\hat{V}^{\bN_i}\}(\xi_0) = \hat{\mathbf{U}}_i\hat{\Lambda}_i\hat{\mathbf{U}}_i^{-1}$.
    \item Define $\hat{f}_i = (\hat{\lambda}_{i1}, ..., \hat{\lambda}_{ik})^{\top}$.
\end{itemize}
\EndFor
\State \Return The vectors $\{\hat{f}_i\}_{i = 1}^n$.
\end{algorithmic}
\end{algorithm}

\begin{algorithm}
\caption{Classification with Fourier-Eigen embeddings.}
\label{alg:classifier}
\begin{algorithmic}
\State \textbf{input}: Data $\{(\hat{f}_i, G_i)\}_{i = 1}^{n}$, function class $\cH$, loss function $L(\cdot,\cdot)$.
\State Solve

\[\hat{h} = \argmin_{h \in \cH}\frac1n\sum_{i = 1}^n L\big(G_i, h(\hat{f}_i)\big).\]
\State \Return The function $\hat{h}$.
\end{algorithmic}
\end{algorithm}

\begin{remark}
    Our proposed embedding procedure in Algorithm \ref{alg:FE} depends on a choice of frequency value $\xi_0$. The choice of $\xi_0$ does not matter in general, since under (\ref{eq:mean-diff}) the vectors $\{\mu^{(g)} : g \in \cG\}$ are assumed to be constant in time and hence are unchanged by taking the Fourier transform. However, we do need the matrix $\mathbf{W}{(\xi_0)}$ to be full rank to ensure that the Fourier-Eigen embeddings corresponding to distinct classes of patients are linearly separated; see Lemma \ref{lemma:gap}. In practice, choosing $\xi_0$ at a moderate frequency yields stable embeddings, while extremely low or high frequencies can be noisier.
\end{remark}

\begin{remark}
Implicitly, Algorithm \ref{alg:FE} assumes knowledge of $k$, the dimension of the latent processes $\{\bM_i\}_{i = 1}^n$. We do not explore estimation of $k$ in the present work, but in practice, users of Algorithm \ref{alg:FE} can specify $k$ with standard methods for estimating the number of nonzero eigenvalues of a matrix \citep{fan2022estimating}.
\end{remark}


\begin{algorithm}
\caption{Clustering with Fourier-Eigen embeddings (Algorithm 1 from \citep{kumar2010clustering}).}
\label{alg:clustering}
\begin{algorithmic}
\State \textbf{input}: Data $\{\hat{f_i}\}_{i = 1}^{n}$.
\State Form the matrix $\bF \in \RR^{n \times k}$, where $\hat{f}_i$ is the $i_{\text{th}}$ row of $\bF$.

\State Let $\hat{\bF}$ denote the rank-$|\cG|$ approximation to $\bF$ computed with the SVD.

\State Calculate initial cluster means $\{\hat{C}_g\}_{g \in \cG}$ as a solution to $|\cG|$-means clustering on the rows of $\hat{\bF}$.

\State Until convergence:
\begin{itemize}
    \item For each $i \in \{1, \ldots, n\}$, set $\hat{G}_i = \argmin_{g \in \cG}\|\hat{f}_i - \hat{C}_g\|_2^2$.
    \item For each $g \in \cG$, set $\hat{C}_g = \frac{1}{|\{i : \hat{G}_i = g \}|}\sum_{i : \hat{G}_i = g}\hat{f}_i$.
\end{itemize}

\State \Return The estimated cluster labels $\{\hat{G_i}\}_{i = 1}^n$.

\end{algorithmic}
\end{algorithm}


\begin{remark}
    Since our embedding method only relies on the cross-covariance matrices of the observed processes $\{\bN_i\}_{i = 1}^n$, we avoid the issue of estimating the baseline intensity functions $\nu_j: \RR^p \rightarrow \RR, j = 1, \ldots, d$. We treat $\nu_j(\bX_i)$ as a nuisance component that absorbs baseline covariate effects and focus on temporal dependence; explicitly incorporating $\bX_i$ into the embedding is an interesting extension left for future work.
\end{remark}

\begin{algorithm}
\caption{Kernel smoothing estimator of cross-covariance matrix.}
\label{alg:cov-est}
\begin{algorithmic}
\State \textbf{input}: Point process $\bN \in \RR^d$ observed from $0$ to $T > 0$, bandwidth $h > 0$, symmetric kernel function $K(\cdot)$, threshold $C_{\text{tr}} > 0$.

\State \Return The function $\tau \mapsto \hat{V}^{\bN_i}(\tau)$, where $\hat{V}^{\bN_i}(\tau)$ has entries

\[\hat{V}^{\bN_i}_{jj'}(\tau) = \begin{cases}
    \frac{1}{Th}\int \int_{[0,T]^2}K\big(\frac{(t' - t) + \tau}{h}\big)dN_{j'}(t')dN_j(t) - \frac{1}{T^2}N_{j'}([0,T])\frac{1}{T^2}N_j([0,T]) & j \neq j' \\
    \frac{1}{Th}\int \int_{[0,T]^2 / \{t=t'\}}K\big(\frac{(t' - t) + \tau}{h}\big)dN_{j'}(t')dN_j(t) - \frac{1}{T^2}\big(N_{j}([0,T])\big)^2 & j = j'
\end{cases},\]
if $|\tau| \leq C_{\text{tr}}$, and $\hat{V}^{\bN_i}_{jj'}(\tau) = 0$ if $|\tau| > C_{\text{tr}}$.
\end{algorithmic}
\end{algorithm}

\section{Theoretical analysis}\label{sec:theory}

In this section, we give a theoretical analysis of our Fourier-Eigen embeddings and their use in downstream classification and classification tasks. 

\begin{assumption}\label{asmp:model}
    The full data $\cD = \{(\bN_i, \bX_i, \bM_i, G_i)\}^n_{i = 1}$ are drawn i.i.d. from  the latent factor point process model (\ref{eq:g-pi} - \ref{eq:lfpp}). For each $i \in [n]$, the processes $M_{i\ell}$ and $M_{i\ell'}$ are independent for $\ell \neq \ell'$. For each $i \in [n]$, the processes $\bN_i$ and $\bM_i$ are stationary. The transfer functions $\{\omega_{j\ell}\}$ are Lebesgue measurable and absolutely integrable. 
\end{assumption}

\begin{lemma}\label{lemma:symmetric}
    Suppose that Assumption \ref{asmp:model} holds. For $g \in \cG$, let $\cV^{(g)}(\xi) = \sF\{V^{\bN_i}\}(\xi)$, where $\bN_i$ is drawn from (\ref{eq:lfpp}) with $G_i = g$. Then at each $\xi \in \RR$, the matrix $\cV^{(g)}(\xi)$ is Hermitian with entries given by
\begin{equation}
\cV_{jj'}^{(g)}(\xi) = \sum_{\ell = 1}^k\mu^{(g)}_{\ell}\cdot \bar{\omega}_{j\ell}(\xi)\cdot \bar{\omega}_{j'\ell}(\xi)
\end{equation}
\end{lemma}

Lemma \ref{lemma:symmetric} guarantees that the population-level Fourier-Eigen embeddings are real-valued, and formalizes our heuristic discussion after Proposition \ref{prop:lowrank}. This allows us to guarantee that the population level Fourier-Eigen embeddings for distinct classes of patients are well-separated in $\ell_2$ norm under the following assumption.


\begin{assumption}\label{asmp:fullrank}
    There exists a frequency value $\xi^{*} \in \RR$ at which the matrix $\mathbf{W}{(\xi^{*})}$ has full column rank.
\end{assumption}


\begin{lemma}\label{lemma:gap}
    Suppose that Assumptions \ref{asmp:model} and \ref{asmp:fullrank} hold. For $g \in \cG$, let $f^{(g)}$ denote the population level Fourier-Eigen embedding obtained at frequency $\xi^*$ for a patient in class $g$ and $\mathbf{U}^{(g)} \in \CC^{d \times k}$ denotes the first $k$ eigenvectors of $\cV^{(g)}(\xi^*)$. Let $g, r \in \cG$ denote an arbitrary pair of classes and define $\rho_{gr} = \|(\mathbf{U}^{(g)})^{\dagger}\mathbf{U}^{(r)} - \mathbf{I}\|_F$. Then the following inequality holds:
    \begin{align}
\frac{1}{k}\|f^{(g)} - f^{(r)}\|_2 &\geq \sigma_k^2(\mathbf{W}) \|\mu^{(g)}- \mu^{(r)} \|_2 - 3\sigma_1^2(\mathbf{W})(\| \mu^{(g)}\|_2 \land \| \mu^{(r)}\|_2) \cdot (\rho_{gr} \lor \rho^2_{gr}).
\end{align}
\end{lemma}

\begin{remark}
    Lemma \ref{lemma:gap} allows us to translate signal between classes in the latent process intensity vectors to differences in the corresponding population-level Fourier-Eigen embeddings. This result formalizes the connection between our latent factor point process model in (\ref{eq:g-pi} - \ref{eq:lfpp}) and our construction of the Fourier-Eigen embeddings as a statistical tool. The inequality stated in the lemma provides a sufficient condition such that $\|f^{(g)} - f^{(r)}\|_2$ is large. The right-hand side is positive for general $\mu^{(g)}$ and $\mu^{(r)}$ as long as the eigenvectors of $\cV^{(g)}(\xi^*)$ and $\cV^{(r)}(\xi^*)$ are well-aligned as captured by the quantity $\rho_{gr}$.  The only identification condition that we need to prove this lemma is Assumption \ref{asmp:fullrank}, which asserts the existence of a frequency $\xi^* \in \RR$ at which $\mathbf{W}(\xi^*)$ is full rank. This is satisfied if, for instance, the transfer functions satisfy $\omega_{j\ell}(t) = a_{j\ell}\kappa(t)$, where $\kappa$ is a Lebesgue measurable function of time and the matrix $\mathbf{A} \in \RR^{d \times k}$ with entries $A_{j\ell} = a_{j\ell}$ is full rank.
\end{remark}

\begin{assumption}\label{asmp:short-tail}
    There exists a constant $b_0 > 0$ such that $\omega_{j\ell}(b) = 0$ for $|b| \geq b_0$ for all $j \in [d]$ and $\ell \in [k]$.
\end{assumption}

\begin{assumption}\label{asmp:bdd-sum}
    There exists a positive constant $\gamma < 1$ such that for all $j \in [d]$,
    \begin{equation}
    \sum_{\ell = 1}^k\int_0^\infty|\omega_{j\ell}(u)|du \leq \gamma.
    \end{equation}
\end{assumption}

\begin{assumption}\label{asmp:cov-lips}
    The elements of the cross-covariance matrix $V^{\bN}$ are $\theta$-Lipschitz functions.
\end{assumption}

\begin{remark}
    Assumptions \ref{asmp:short-tail} - \ref{asmp:cov-lips} are required to apply the concentration properties of the kernel-smoothing estimator of the cross-covariance matrix given in Algorithm \ref{alg:cov-est}. These are standard in the analysis of high-dimensional point processes; see for instance Assumptions 2 and 3 of \citet{chen2017multivariate}. We remark that our Assumption \ref{asmp:short-tail} is strictly stronger than Assumption 2 of \citet{chen2017multivariate}, as we require that the tails of the transfer functions $\omega_{j\ell}$ are identically 0 at long time horizons. This is used to control the fluctuations of the Fourier-transformed cross-covariance matrix; see the appendix for details.
\end{remark}

We now introduce the following notation to present our first theorem, which states a bound on the generalization error of a classifier trained on the estimated Fourier-Eigen embeddings with Algorithm \ref{alg:classifier} over the function class $\cH$. Let $\rho_\cG(\cdot, \cdot)$ denote a real-valued dissimilarity measure on the set of classes $\cG$; for instance, $\rho_\cG(g, r) = \one(g \neq r)$. For a constant $\lambda_{\text{Lip}} > 0$, we say that the loss function $L(\cdot, \cdot)$ is $\lambda_{\text{Lip}}$-Lipschitz if, for any fixed $g \in \cG$, it holds
\begin{equation}
|L(g,r) - L(g,r')| \leq \lambda_{\text{Lip}}\rho_\cG(r, r'),
\end{equation}
for $r, r' \in \cG$. We define the diameter of the set $\cG$ with respect to $\rho_\cG$ as
\begin{equation}
    \text{diam}(\cG) = \sup_{g, r \in \cG}\rho_\cG(g, r).
\end{equation}
Additionally, we define the oracle classifier $h_{\text{oracle}} : \RR^k \rightarrow \cG$ as follows:
\begin{equation}
h_{\text{oracle}}(x) = \argmin_{g \in \cG}\|f^{(g)} - x\|_2^2.\end{equation}
Finally, we define
\begin{equation}\cE_{\text{approx.}}\big(\cH\big) = \inf_{h \in \cH}\sup_{x \in \RR^k}\rho_\cG\big(h(x), h_{\text{oracle}}(x)\big),
\end{equation}
as the approximation error of the function class $\cH$. We will measure the complexity of $\cH$ in terms of its Rademacher complexity \citep{bartlett2002rademacher} under composition with the loss function $L(\cdot, \cdot)$, which is defined as
\begin{equation}
    \cR(L \circ \cH) = \EE \sup_{h \in \cH}\Big|\frac1n \sum_{i = 1}^n\varepsilon_i L\big(G_i, h(\hat{f}_i)\big)\Big|,
\end{equation}
where $\varepsilon_1, \ldots, \varepsilon_n$ are i.i.d. Rademacher random variables, and the pairs $\{(G_i, \hat{f}_i)\}_{i = 1}^n$ constitute the training data passed to Algorithm \ref{alg:classifier}. The outer expectation is taken jointly over the Rademacher random variables and the training data.

\begin{theorem}\label{thm:classification}
    Suppose that Assumptions \ref{asmp:model}, \ref{asmp:fullrank}, \ref{asmp:short-tail}, \ref{asmp:bdd-sum}, and \ref{asmp:cov-lips} hold. Given the data $\cO_{\text{sup}} = \{(\bN_i, \bX_i, G_i)\}_{i = 1}^n$,, let $\{\hat{f}_i\}_{i = 1}^n$ denote the estimated Fourier-Eigen embeddings returned from Algorithm \ref{alg:FE} at frequency $\xi^*$, where Algorithm \ref{alg:cov-est} is implemented with bandwidth $h = c_1T_i^{-1/5}$ for each $i \in \{1, \ldots, n\}$ and threshold $C_{\text{tr}} \leq 2b_0$ for a constant $c_1 > 0$. Suppose that there exists a constant $\lambda_{\text{Lip}} > 0$ such that $L(\cdot, \cdot)$ is $\lambda_{\text{Lip}}$-Lipschitz. Let $\hat{h}$ denote the classifier trained on the pairs $\{(\hat{f}_i, G_i)\}_{i = 1}^n$ with Algorithm \ref{alg:classifier} using the function class $\cH$ and loss function $L(\cdot, \cdot)$ and let $(\hat{f}, G)$ denote a new pair independently drawn from the same distribution as the training data with observation time $T > 0$. Then there exist constants $c_2, c_3, C > 0$ such that if for any pair of classes $g, r \in \cG$, $\mu^{(g)}$ and $\mu^{(r)}$ satisfy
    \begin{equation}\label{eq:classification-ss}
    \|\mu^{(g)} - \mu^{(r)}\|_2 \geq C\frac{d}{\sqrt{k}\sigma_k^2(\mathbf{W})}T^{-1/5} + 3\frac{\sigma_1^2(\mathbf{W})}{k}(\| \mu^{(g)}\|_2 \land \| \mu^{(r)}\|_2) \cdot (\rho_{gr} \lor \rho^2_{gr})\big),\end{equation}
    then it holds
    \begin{equation}\label{eq:classification-bound}
        \EE\Big[L\big(G, \hat{h}(\hat{f})\big)\Big] \leq \lambda_{\text{Lip}}\cE_{\text{approx.}}(\cH) + 2c_2 d^2 T^{6/5}\exp\big(c_3 T^{-1/5}\big) \cdot \text{diam}(\cG) + 4 \cdot  \cR(L \circ \cH).
    \end{equation}
\end{theorem}

The conclusion of Theorem \ref{thm:classification} takes the form of standard risk bounds for empirical risk minimizers \citep{vapnik2013nature, bartlett2002rademacher, hajek2021ece}. The first term on the right hand side of (\ref{eq:classification-bound}) encodes the ability of functions in $\cH$ to approximate an oracle classifier that relies on knowledge of the population-level Fourier-Eigen embeddings, the second term denotes the worst-case error of the oracle classifier, and the final term accounts for the complexity of the function class $\cH$. We can refine this result in the case of binary classification, as given in the following corollary. Here, we let $\text{VC}(\cH)$ denote the Vapnik-Chervonenkis dimension \citep{devroye1996vapnik} of $\cH$.

\begin{corollary}\label{cor:binary-classification}
    Suppose that all of the conditions of Theorem \ref{thm:classification} hold. Furthermore, suppose that $\cG = \{0,1\}$ and $\rho_\cG(g, r) = |g - r|$ for $g, r \in \cG$. Then there exist constants $c_2, c_3, C, C' > 0$ such that if (\ref{eq:classification-ss}) holds for any pair of classes $g, r \in \cG$, then
    \begin{equation}
        \EE\Big[L\big(G, \hat{h}(\hat{f})\big)\Big] \leq \lambda_{\text{Lip}}\cE_{\text{approx.}}(\cH) + 2c_2 d^2 T^{6/5}\exp\big(c_3 T^{-1/5}\big) + C' \sqrt{\frac{\text{VC}(\cH)}{n}}.
    \end{equation}
\end{corollary}

\begin{remark}
    Corollary \ref{cor:binary-classification} allows us to evaluate the statistical complexity of binary classification for standard choices of the function class $\cH$. For instance, it is well known that the class of linear classifiers $\cH_{\text{lin}}$ on $\RR^k$ satisfies $\text{VC}(\cH_{\text{lin}}) \asymp k$ \citep{hajek2021ece}, and the class of feedforward neural networks $\cH_{\text{FF}}$ with $w$ weights has $\text{VC}(\cH_{\text{FF}}) \lesssim  w \log w$ \citep{baum1988size}. This also emphasizes one of the advantages of using our Fourier-Eigen embeddings for classification rather than the raw EHR trajectory data, as the reduction in dimension granted by our embeddings can significantly reduce the size of $\text{VC}(\cH)$, especially if $k \ll d$.
\end{remark}

We now turn our attention to the performance of Algorithm \ref{alg:clustering}. In the following theorem, we let $n_g = \sum_{i = 1}^n \one(G_i = g)$ for each $g \in \cG$ and we define $\cS_{\cG}$ as the set of permutations on $\cG$, meaning that $\sigma \in \cS_{\cG}$ implies that $\sigma$ is a bijection from $\cG$ to $\cG$.
\begin{theorem}\label{thm:clustering}
    Suppose that Assumptions \ref{asmp:model}, \ref{asmp:fullrank}, \ref{asmp:short-tail}, \ref{asmp:bdd-sum}, and \ref{asmp:cov-lips} hold. Given the data $\cO_{\text{unsup}} = \{(\bN_i, \bX_i)\}_{i = 1}^n$, let $\{\hat{f}_i\}_{i = 1}^n$ denote the estimated Fourier-Eigen embeddings returned from Algorithm \ref{alg:FE} at frequency $\xi^*$, where Algorithm \ref{alg:cov-est} is implemented with bandwidth $h = c_1T_i^{-1/5}$ for each $i \in \{1, \ldots, n\}$ and threshold $C_{\text{tr}} \leq 2b_0$ for a constant $c_1 > 0$. Finally, let $\{\hat{G}_i\}_{i = 1}^n$ denote the estimated cluster labels obtained from Algorithm \ref{alg:clustering} and $T_{\min} = \min_{i = 1}^n T_i$. Then the following claim holds conditional on the labels $\{ G_i \}_{i = 1}^n$: there exist constants $c_2, c_3, C > 0$ such that if for any pair of classes $g, r \in \cG$, $\mu^{(g)}$ and $\mu^{(r)}$ satisfy
    \begin{equation}\label{eq:clustering-ss} \|\mu^{(g)} - \mu^{(r)}\|_2 \geq C\frac{d |\cG|}{\sigma_k^2(\mathbf{W})}\sqrt{\frac{n}{k}}\left(\frac{1}{\sqrt{n_g}} + \frac{1}{\sqrt{n_r}}\right)T_{\min}^{-1/5} + 3\frac{\sigma_1^2(\mathbf{W})}{k}(\| \mu^{(g)}\|_2 \land \| \mu^{(r)}\|_2) \cdot (\rho_{gr} \lor \rho^2_{gr})\big),\end{equation}
    then it holds
    \begin{equation}\inf_{\sigma \in \cS_{\cG}}\left[\sum_{i = 1}^n\one (G_i \neq \sigma(\hat{G}_i))\right] = 0\end{equation}
    with probability at least $1 - 2c_2nd^2 T_{\min}^{6/5}\exp\left(-c_3 T_{\min}^{1/5}\right)$.
\end{theorem}
Theorem \ref{thm:clustering} shows that a standard K-means clustering algorithm can perfectly recover the latent group membership of each patient, as long as the signal separating the groups is sufficiently strong. We point out that the signal strength condition (\ref{eq:clustering-ss}) is stronger than (\ref{eq:classification-ss}) as required in Theorem \ref{thm:classification}. We interpret this gap as the ``cost'' of unsupervised learning, which is minimal when we observe patients from a small number of balanced classes.

\section{Simulation studies}\label{sec:sims}

We conduct simulation studies to evaluate the performance of classification and clustering with our proposed Fourier-Eigen embeddings. We simulate data $\{\bN_i\}_{i = 1}^{n}$ from our latent factor point process model using the thinning algorithm \citep{daley2003introduction} with two classes of patients of equal size. We fix the ambient dimension $d = 100$ and the latent dimension $k = 2$, and vary the total sample size $n \in \{250,500,750\}$ and observation time $T \in \{50, 100, 150\}$. We specify the transfer functions as $\omega_{j\ell}(t) = a_{j\ell}e^{-t^2/2}$, where the coefficients $a_{j\ell}$ are drawn from a $\text{Unif}(0,1/2)$ distribution. We vary the signal strength between the two groups, as encoded by the difference between the latent intensity vectors $\mu^{(0)}$ and $\mu^{(1)}$, as $\|\mu^{(0)} - \mu^{(1)}\|_2 \in \{0.3, 0.4, 0.5, 0.6, 0.7, 0.8\}$.

\begin{figure}[h!] 
     \centering
     \includegraphics[width=1\textwidth]{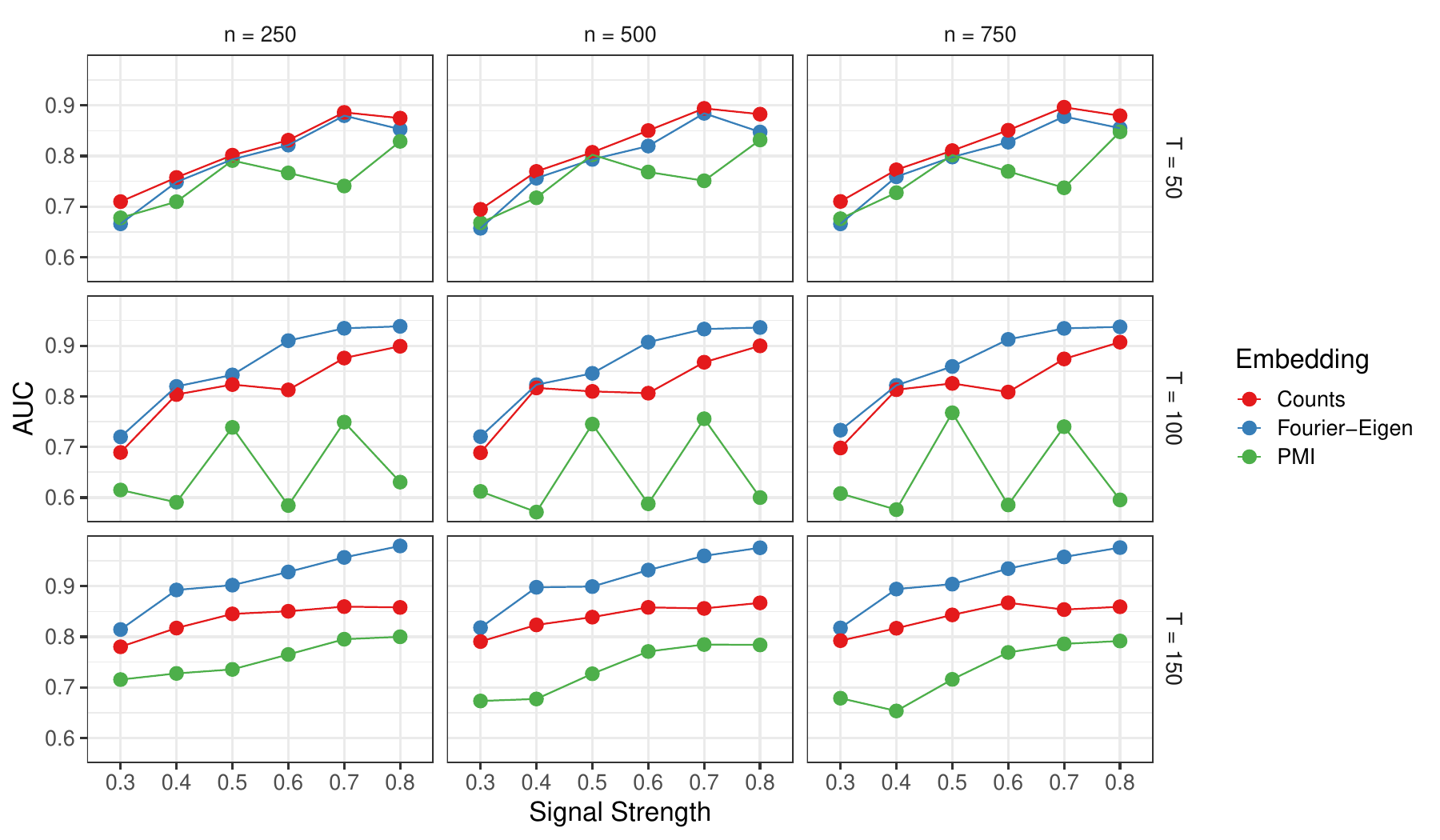} 
     \caption{Average AUC over 100 replications. The $x$ axis denotes the signal strength $\|\mu^{(0)} - \mu^{(1)}\|_2$.}
     \label{fig:auc}
 \end{figure}

We compute our Fourier-Eigen embeddings with Algorithm \ref{alg:FE} at frequency $\xi_0 = 1$, where the cross-covariance matrix is estimated with Algorithm \ref{alg:cov-est} using bandwidth $h = 1$, $C_{\text{tr}} = 5$, and a Gaussian kernel function. We calculate the Fourier transform of the cross-covariance matrix by numerical integration. We compare the performance of the Fourier-Eigen embeddings on downstream tasks to the follow two embedding methods:
\begin{enumerate}
    \item \underline{Code counts:} For each $i \in \{1, \ldots, n\}$, we compute a feature vector in $\RR^d$ by summing the number of occurrences of each code.
    \item \underline{Point-wise mutual information (PMI) embeddings:} We calculate the PMI matrix for each simulated process $N_i$ according to the formulation in \citet{arora2016latent} with window size $T / 20$. We use the first $k = 2$ eigenvalues of the PMI matrix as the feature vector. 
\end{enumerate}
First we evaluate the performance of each of these three embedding methods for classification. We train three sets of standard logistic regression models using the class membership indicators as outcomes and the Fourier-Eigen embeddings, code counts, and PMI embeddings as features respectively. We evaluate the prediction performance of each embedding via out-of-sample AUC on a testing set of size $n_\text{test} = 50$. The results are given for each configuration of $n$, $T$, and signal strength in Figure \ref{fig:auc}. The AUC is averaged over 100 replications. 

We can see that the discriminative performance of Fourier-Eigen embeddings becomes especially apparent as the observation time $T$ increases. At $T = 50$, the three feature embeddings admit very similar classification performance, with the code counts and Fourier-Eigen embeddings slightly outperforming the PMI embeddings. However, the logistic regression models trained on the PMI embeddings and code counts do not improve with increasing $T$, as these methods are unable to take advantage of increased temporal information due to model misspecification. Our Fourier-Eigen embeddings increasingly out-perform the competing methods for $T \in \{100,150\}$, since the estimated embedding vectors concentrate more sharply around the population-level embeddings which are provably well-separated due to Lemma \ref{lemma:gap}.

\begin{figure}[h!] 
     \centering
     \includegraphics[width=1\textwidth]{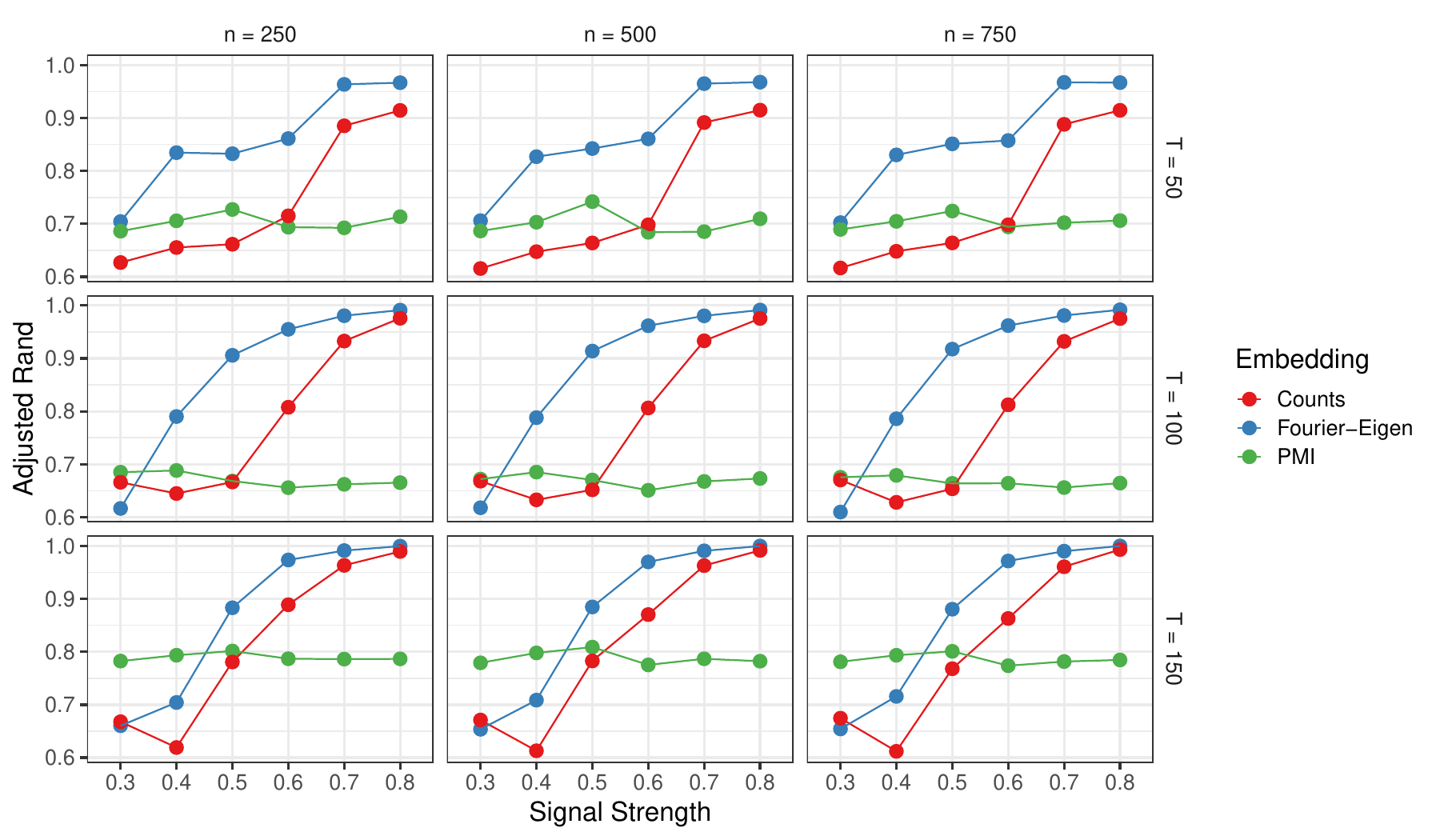} 
     \caption{Average adjusted Rand index over 100 replications. The $x$ axis denotes the signal strength $\|\mu^{(0)} - \mu^{(1)}\|_2$.}
     \label{fig:rand}
 \end{figure}

Next, we run K-means clustering on each set of embeddings with $2$ clusters. We evaluate the performance of each method by the adjusted Rand index, using the true class membership indicators as our reference labels. The results for each configuration of $n$, $T$, and signal strength are presented in Figure \ref{fig:rand}, averaged over 100 replications. Here, we see that our method is able to recover the true class membership labels with greater accuracy than either the code counts or PMI embeddings, especially as the signal strength increases. The Fourier-Eigen embeddings perform poorly when the signal between the classes is too weak, which is consistent with Theorem \ref{thm:clustering}. As the signal strength increases, the Fourier-Eigen embeddings recover the true class labels more efficiently than either of the competing methods, with the count-based clustering results aligning with the Fourier-Eigen results in the very high signal setting. The PMI embeddings return lackluster clustering performance across all settings, which we attribute to the restrictive assumption of a log-linear intensity function that PMI-based methods enforce on the code occurrence probabilities.

 Finally, we evaluate the robustness of our Fourier-Eigen embeddings to the functional form of the transfer functions $\{\omega_{j\ell}\}$. We set $\omega_{j\ell}(t) = a_{j\ell}\beta(t)$ for $a_{j\ell} \sim \text{Unif}(0,1/2)$ and we vary the transfer kernel function $\beta(t)$ in the set $\beta(t) \in \big\{\frac{|\sin(t)|_{\one(t < \pi)}}{t+1},(1 - \sqrt{t})_{\one(t < 1)}, (1 - t)_{\one(t < 1)}, 4^{-t}_{\one(t < 2)}\big\}$. We simulate data from two groups with signal strength $\|\mu^{(0)} - \mu^{(1)}\|_2 = 1.6$, ambient dimension $d = 100$, latent dimension $k = 2$, observation time $T = 100$, and total sample size $n = 500$. For each choice of $\beta(t)$, we compare the efficacy our Fourier-Eigen embeddings to the PMI embeddings and code counts in classification with logistic regression and clustering, again measuring the performance of each method with AUC and the adjusted Rand index. The results are given (averaged over 100 replications) in Table \ref{tab:diffomega-results}. 
 
 The Fourier-Eigen embeddings give the best performance for a majority of the choices of $\beta(t)$ in terms of AUC. For the clustering task, our method gives the highest adjusted Rand index in half of the cases, with the count-based clustering achieving the best performance for $\beta(t) = (1 - \sqrt{t})_{\one(t < 1)}$ and $\beta(t) = (1 - t)_{\one(t < 1)}$. However, the Fourier-Eigen embeddings perform nearly as well as the counts in both of these settings, resulting in an average adjusted Rand score that is far greater than the competing methods. These results indicate that our Fourier-Eigen embeddings can provide strong statistical perform in downstream tasks across a variety of forms for the conditional intensity function of the observed data, as suggested by our theoretical results.

\begin{table}[h!]
\centering
 \begin{tabular}{|c|c|cccc|c|}
    \hline
    Metric &
    Embedding & $ \frac{|\sin(t)|_{\one(t < \pi)}}{t+1}$ & $(1 - \sqrt{t})_{\one(t < 1)}$ & $(1 - t)_{\one(t < 1)}$ & $4^{-t}_{\one(t < 2)}$ & Average \\
    \hline
    \multirow{3}{*}{AUC} & Fourier-Eigen & $\mathbf{0.8868}$ & $\mathbf{0.7406}$ & $\mathbf{0.6645}$ & 0.7256 & $\mathbf{0.7544}$ \\
     &
    PMI  & 0.8763 & 0.6654 & 0.6252 & $\mathbf{0.8126}$& 0.7449 \\
    & Counts & 0.8866 & 0.5547 & 0.5650 & 0.6836 & 0.6725\\
    \hline
    \multirow{3}{*}{Adj. Rand} & Fourier-Eigen & $\mathbf{0.9606}$ & 0.8595 & 0.9364 & $\mathbf{0.8911}$ & $\mathbf{0.9119}$ \\
     &
    PMI  & 0.9467 & 0.7935 & 0.7764 & 0.6640 & 0.7952\\
    & Counts & 0.8348 & $\mathbf{0.9389}$ & $\mathbf{0.9432}$ & 0.6777 & 0.8487\\
    \hline
\end{tabular}
\caption{Average AUC and adjusted Rand index over 100 replications for varying specifications of the transfer kernel function $\beta(t)$. The score of the best performing method for each $\beta(t)$ is rendered in bold. The final column gives the average score across all four specifications of $\beta(t)$.}\label{tab:diffomega-results}
\end{table}

\section{Real data analysis: Alzheimer's disease severity}

We demonstrate the real-world utility of our proposed Fourier-Eigen embedding method on an analysis of electronic health records data from the University of Pittsburgh Medical Center (UPMC). Our objective is to cluster patients with Alzheimer's disease (AD) according to severity of AD progression using EHR code occurrence data. Alzheimer's disease is widely regarded to be the most common cause of dementia, which affects over 50 million people worldwide and leads to severe  declines in quality of life and increased rates of mortality \citep{korczyn2024alzheimer}. As the rate of progression of AD has consequences for long-term health outcomes \citep{tahami2022alzheimer}, there is a great interest in leveraging the temporal information in EHR patient trajectories to predict progression patterns and identify at-risk patient groups \citep{fu2025identifying, kumar2021machine, akter2025using, li2023early, venkatesh2025leveraging}. In this analysis, we aim to evaluate the ability of our Fourier-Eigen embeddings to distinguish patients with different AD progression patterns, in comparison to competing methods. We use nursing home admission time as a proxy for AD progression to compensate for a lack of gold-standard labels from chart review.


Our dataset consists of 92,723 electronic health records of patients who have at least one occurrence of an AD diagnosis code (PheCode:290.11), including 48,454 patients who were admitted to a nursing home and 44,269 who were not. We define the observation period for each patient as the two years prior to the date of the first occurrence of an AD-related code. We remove codes that were present in only $<5\%$ of individuals, which leaves us with $d = 1,375$ codes, including 545 disease diagnosis codes (PheCode), 80 procedure codes (CCS), 273 prescription codes, 262 laboratory codes (LOINC), and 215 local codes that do not belong to any of these four categories. From here, we draw a random subset of $8000$ patients for the analysis, split evenly between patients with a recorded nursing home admission time and patients who do not. Using each patient's trajectory, we compute the following three embeddings: 1) Fourier-Eigen embeddings of dimension $k = 5$ using a Gaussian kernel with a half-week bandwidth, and frequency $\xi_0 = 1$;  2) PMI embeddings of dimension $k = 5$ as defined in Section \ref{sec:sims}, using a co-occurrence matrix computed with a week-long window size; 3) BERT-based embeddings \citep{devlin2019bert}. Specifically, we compute the BERT embeddings by pulling a dimension $m = 769$ representation of each code from a pretrained BERT model, giving us the code-representation matrix $\mathbf{C} \in \RR^{d \times m}$. Letting $\mathbf{Z} \in \RR^{n \times d}$ denote the code count matrix (defined in Section \ref{sec:sims}), we define the BERT embeddings as the first $k = 5$ left singular vectors of the matrix $\mathbf{Z}\mathbf{C}$. 

\begin{figure}[h!] 
     \centering
     \includegraphics[width=\textwidth]{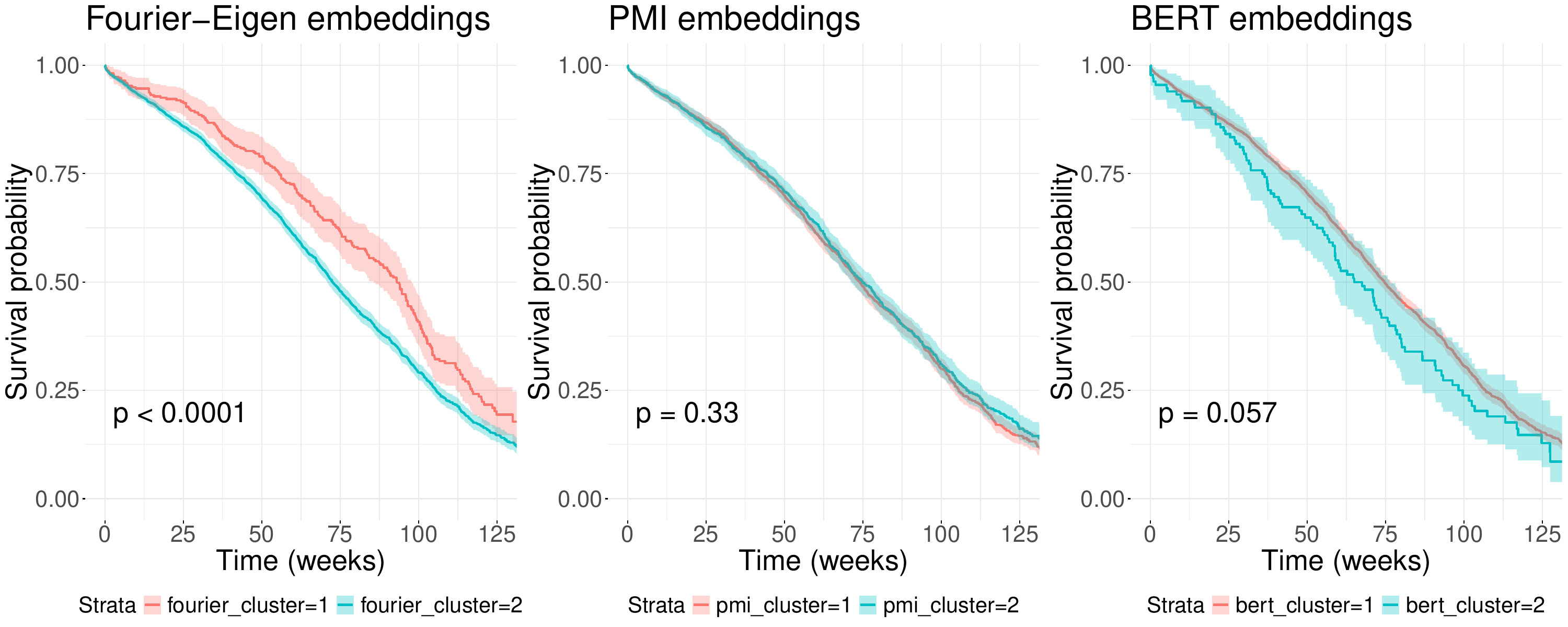} 
     \caption{Kaplan-Meier curves for time of nursing home admission, colored by the cluster assigned to each patient. The figures correspond to the features that were used as input to the clustering algorithm.}
     \label{fig:km-grid}
 \end{figure}

With each set of embeddings as our matrix of features, we run K-means clustering with two clusters to obtain an assigned subgroup label for each patient. To evaluate the ability of these cluster assignments to discern subgroups with differing AD progression, we compute a Kaplan-Meier curve within each cluster, using time in weeks from the first code occurrence to nursing home admission as the outcome. For patients who were not admitted to a nursing home, we define the right-censoring time as the final code occurrence time in their trajectory. The Kaplan-Meier curves with a 95\% confidence band and the p-value from a standard log-rank test are given in Figure \ref{fig:km-grid}. It is clear that our Fourier-Eigen embedding approach is the only embedding method able to discern two significant groups of patients, correspond to fast and slow Alzheimer's progression. The BERT-based embeddings produce two clusters with different Kaplan-Meier curves, but the log-rank test is not significant due to high variance. Meanwhile, the PMI embeddings are not able to discern any signal in the data, likely due to severe model misspecification.

We supplement these results with a Cox regression analysis. For each of the three embedding methods, we fit a standard Cox model using time to nursing home admission as the outcome and the clustering membership indicator as the covariate. In Table \ref{tab:cox-results}, we provide the estimated hazard ratios, the 95\% confidence intervals, and p-values from a likelihood ratio test for each of the three Cox models. The model fit on the clusters produced by the Fourier-Eigen embeddings is the only one able to detect a significant hazard ratio between the subgroups. These results strongly suggest that the Fourier-Eigen embeddings are able to encapsulate interpretable signal between the EHR trajectories from heterogeneous groups of patients, while the PMI and BERT embeddings are clearly less powerful.

\begin{table}[h!]
\centering
 \begin{tabular}{|c|cccc|}
    \hline
    \textbf{Embedding} & \textbf{Estimated HR} & \textbf{Lower 95\% CI} & \textbf{Upper 95\% CI} & \textbf{p-value} \\
    \hline
    Fourier-Eigen & 1.344 & 1.17 & 1.545 & 3.04e-05$^{***}$ \\
    PMI  & 0.9586 & 0.8805 & 1.044 & 0.33 \\
    BERT & 0.8237 & 0.6741 & 1.006 & 0.07 \\
    \hline
\end{tabular}
\caption{Results of Cox regression analysis with cluster membership indicators.}\label{tab:cox-results}
\end{table}

\begin{figure}[h!] 
    \centering 

    \begin{minipage}{0.45\textwidth} 
        \centering
        \includegraphics[width=\textwidth]{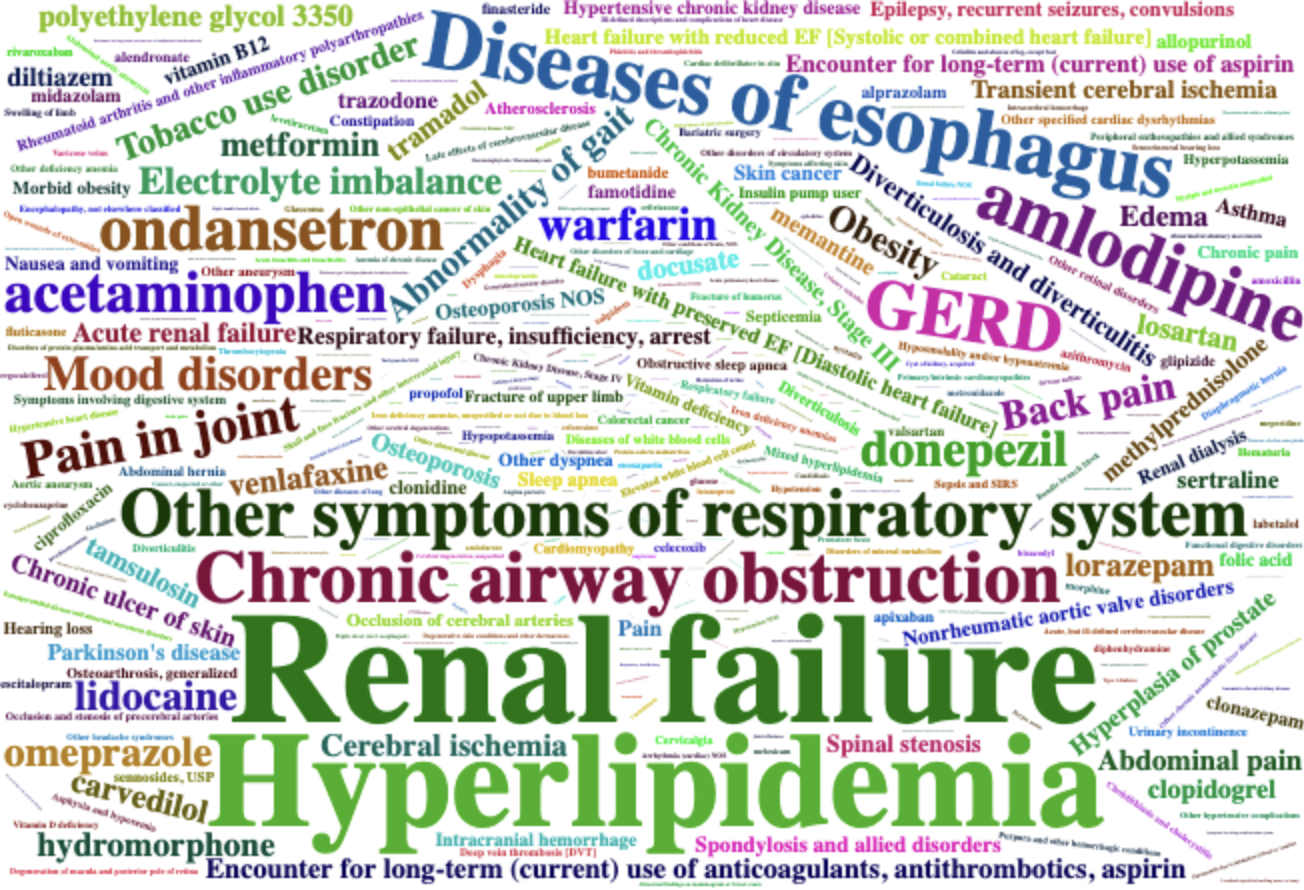}
        \subcaption{Slow progression cluster.}
        \label{fig:wc1}
    \end{minipage}
    \hfill 
    \begin{minipage}{0.45\textwidth} 
        \centering
        \includegraphics[width=\textwidth]{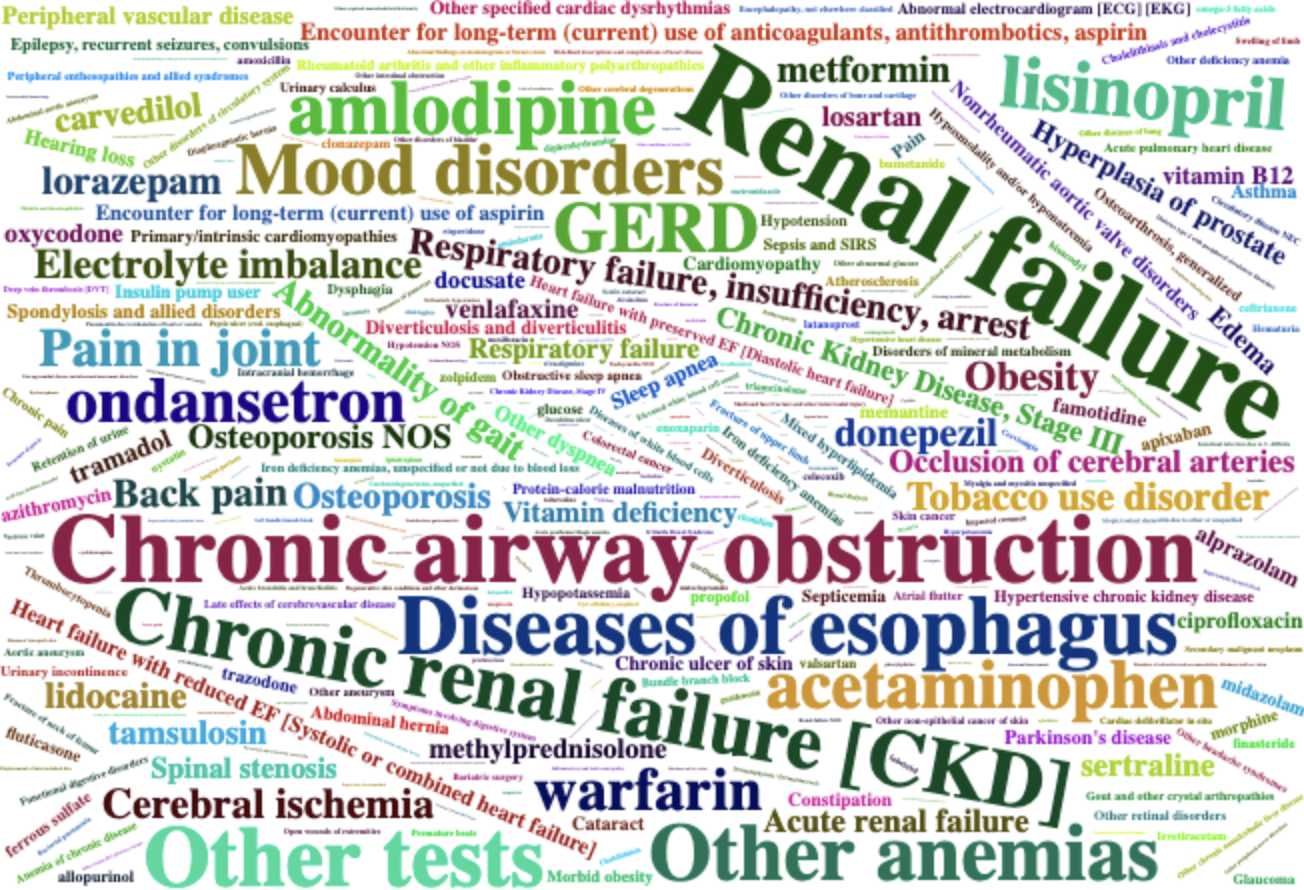}
        \subcaption{Fast progression cluster.}
        \label{fig:wc2}
    \end{minipage}

    \caption{Word clouds of EHR code descriptions for the slow and fast progression clusters obtained from Fourier-Eigen embeddings. The size of each word is determined by its relative frequency within each cluster.}
    \label{fig:worldclouds}
\end{figure}

Finally, we perform an exploration of the most prevalent EHR features in each of the two clusters derived from our Fourier-Eigen embeddings. In Figure \ref{fig:worldclouds}, we provide word clouds of the EHR code descriptions, sized by relative within-cluster prevalence, for the slow and fast progression clusters. We can see that the fast progression cluster admits higher rates of chronic airway obstruction and chronic renal failure, both of which have been found to be associated with severe Alzheimer's disease and dementia \citep{stanciu2020renal,zhang2020association, tondo2018chronic}. Meanwhile, the slow progression cluster seems to have a relatively higher number of occurrences of non-chronic renal failure, suggesting that the Fourier-Eigen embeddings identify a cluster of patients with more severe kidney disease, leading to faster Alzheimer's progression. The fast progression cluster also presents a higher prevalence of mood disorders which is well known to be associated with Alzheimer's risk \citep{ownby2006depression} and anemias, which may complicate AD progression by increasing risk of cognitive decline \citep{faux2014anemia}.

\section{Conclusion and discussion}

In this work, we have described a new model for high-dimensional point processes, called the \textit{latent factor point process} model, whose structure is inspired by electronic health records data. We have shown that interpretable differences between subgroups of patients can be described with a spectral analysis of the cross-covariance matrix under this model, which we leverage to define low-dimensional patient embeddings called \textit{Fourier-Eigen embeddings}. We have demonstrated through theoretical results, simulations, and real data analysis that our Fourier-Eigen embeddings capture signal that can be used to build classification models or accurately cluster patients that belong to different subgroups.

Our contributions are not without limitations, and we foresee several promising directions for future research. First of all, we have imposed that observed and latent processes $(\bN_i, \bM_i)$ are stationary for all $i \in \{1, \ldots, n\}$. This excludes cases where the intensity of the latent process may change over time, such as when a patient changes from one class to another (for instance, ``healthy'' to ``sick''). Relaxing this assumption would greatly improve the real-world applicability of our Fourier-Eigen embeddings, and would open up new use-cases such as changepoint detection and localization. Additionally, our model only allows for excitatory effects, as we assume that the transfer functions satisfy $\omega_{j\ell}(t) \geq 0$ for all $j, \ell$ and all $t \geq 0$. Allowing for latent processes that decrease the probability of a particular code occurrence is of great interest, and extensions in this direction could build on recent work studying point processes with inhibitory effects \citep{chen2017multivariate, gao2024learning, pfaffelhuber2022mean, bonnet2021maximum, cai2024latent}.


\bibliographystyle{ims}
\bibliography{ref}

\newpage


\end{document}